\documentclass[%
twocolumn,
nofootinbib,
amsmath,
amssymb,
aps,
pre
]{revtex4-2}

\usepackage{graphicx,xcolor}% Include figure files
\usepackage{dcolumn}% Align table columns on decimal point
\usepackage{bm}% bold math
\usepackage{graphicx}
\usepackage{wrapfig}
\usepackage{float}
\usepackage{pgfplots}
\pgfplotsset{width=10cm,compat=1.9}

\newcommand{\treetilde}{}

\begin{document}

\preprint{APS/123-QED}

%%%
\title{Configurational entropy of randomly double-folding ring polymers}

%
\author{Pieter H. W. van der Hoek}
\email{pvanderh@sissa.it}
\affiliation{
SISSA - Scuola Internazionale Superiore di Studi Avanzati, Via Bonomea 265, 34136 Trieste, Italy
}
\author{Angelo Rosa}
\email{anrosa@sissa.it}
\affiliation{
SISSA - Scuola Internazionale Superiore di Studi Avanzati, Via Bonomea 265, 34136 Trieste, Italy
}
\author{Elham Ghobadpour}
\email{elham.ghobadpour@ens-lyon.fr}
\affiliation{
ENS de Lyon, CNRS, Laboratoire de Physique (LPENSL UMR5672) et Centre Blaise Pascal, 69342 Lyon cedex 07, France
}
\author{Ralf Everaers}
\email{ralf.everaers@ens-lyon.fr}
\affiliation{
ENS de Lyon, CNRS, Laboratoire de Physique (LPENSL UMR5672) et Centre Blaise Pascal, 69342 Lyon cedex 07, France
}

\date{\today}% It is always \today, today,

%%%
\begin{abstract}
%%%
Topologically constrained genome-like polymers often double-fold into tree-like configurations.
Here we calculate the exact number of tightly double-folded configurations available to a ring polymer in ideal conditions.
For this purpose, we introduce a scheme which allows us to define a ``code'' specifying how a ring wraps a randomly branching tree and calculate the number of admissible wrapping codes via a variant of Bertrand's ballot theorem.
As a validation, we demonstrate that data from Monte Carlo simulations of an elastic lattice model of non-interacting tightly double-folded rings with controlled branching activity are in excellent agreement with exact expressions for branch-node and tree size statistics that can be derived from our expression for the ring entropy.
%%%
\end{abstract}
%%%

\maketitle

%\tableofcontents

\begin{figure}%[ht]
%\centering
\includegraphics[width=0.48\textwidth]{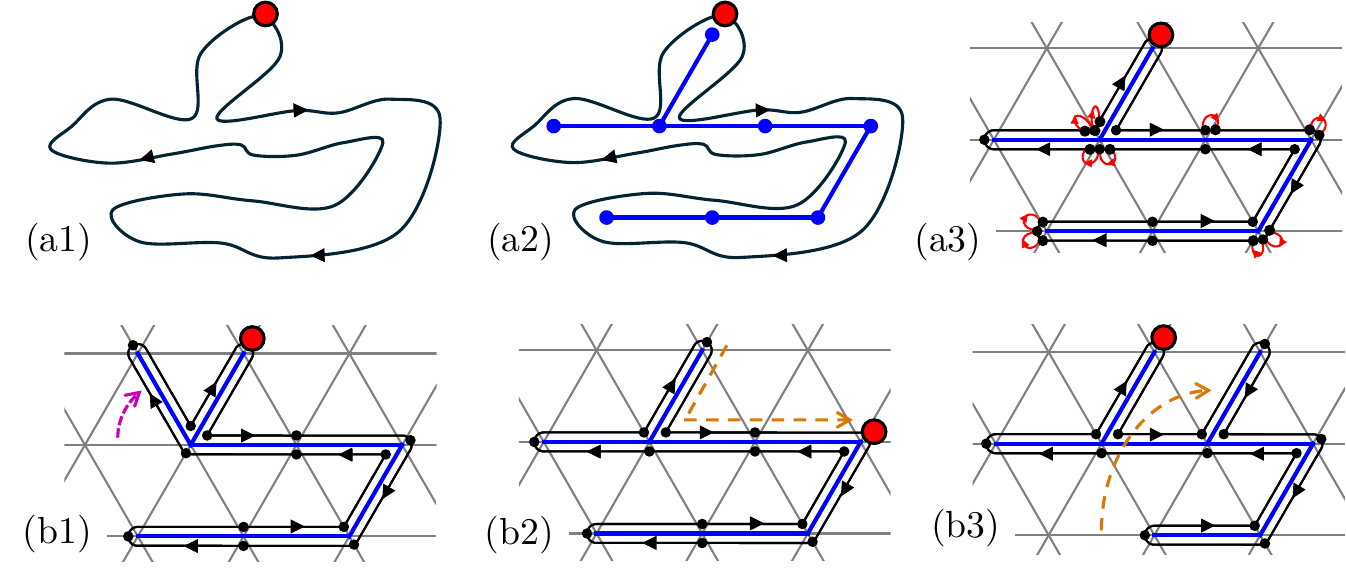}
% Answer: [trim={left bottom right top},clip]
% Ex. 1: trim from left edge
\caption{
Illustration of the modeling steps and notation used in this work.
In various physical and biological situations, (ring) polymers double-fold into conformations (a1) that can be characterised via acyclic trees (a2) and represented by (elastic) lattice models (a3) along the lines of Refs.~\cite{RubinsteinPRL1994,Ghobadpour2021,Ghobadpour2025} where we have highlighted the monomer labeled “1” in red.
(b1) Another {\em conformation} of the embedded ring for the same {\em configuration} of the tree and the same secondary structure of the double-folded ring.
(b2) A cyclic permutation of the ring around the tree corresponds to a different {\em configuration} or secondary structure of the ring.
(b3) A different tree {\em configuration} and hence also a different {\em configuration} of the double-folded ring.
}
\label{fig:Modelfigure}
\end{figure}

%%%
{\it Introduction.}
%%%
Topologically constrained genome-like polymers often double-fold into tree-like configurations
as they form plectonemes due to supercoiling~\cite{MarkoSiggia1994,MarkoSiggiaSuperCoiledDNA1995,Woldringh1999,Cunha2001},
undergo loop extrusion~\cite{AlipourMarkoNAR2012,Sanborn2015,Fudenberg2016,GoloborodkoBJ2016,GoloborodkoELife2016},
or maximize the entropy of the crumpled~\cite{grosbergEPL1993} territorial~\cite{CremerReview2001} arrangement of interphase chromosomes arising from the decondensation of topologically untangled metaphase chromosomes~\cite{RosaPLOS2008,Dekker-Hic2009,RosaBJ2010}.
Randomly branching double-folded ring polymers were first  discussed theoretically~\cite{KhokhlovNechaev85,RubinsteinPRL1986,RubinsteinPRL1994,GrosbergSoftMatter2014,SmrekGrosberg2015} and then explored numerically~\cite{Halverson2011_1,MichielettoSoftMatter2016,RosaEveraersPRL2014,SchramSM2019,SmrekRosa2019,UbertiniRosa2025} in the context of dense solutions and melts of unknotted and non-concatenated ring polymers.
Fig.~\ref{fig:Modelfigure}(a) schematizes the progression from an off-lattice representation of a double-folded ring, over the association of a randomly branching tree characterising the secondary structure~\cite{RosaEveraers2019}, to a lattice model~\cite{RubinsteinPRL1994,Ghobadpour2021,Ghobadpour2025} of tightly double-folded rings with reptons~\cite{RubinsteinRepton1987} representing stored length.
The purpose of the present work is to calculate the configurational entropy of randomly double-folded rings in ideal conditions. 
To this end, we introduce a scheme which allows us to define a ``code'' specifying how a ring wraps a randomly branching tree, calculate the number of admissible wrapping codes, and present a detailed comparison of theoretical predictions and corresponding simulation data for a variant of the elastic lattice model introduced in Refs.~\cite{Ghobadpour2021,Ghobadpour2025} which allows us to control the branching activity.

%%%
%%%

%%%
\begin{figure*}%[ht]
%\centering
\includegraphics[page=2,width=0.75\textwidth,trim={3.2cm 4.3cm 2.7cm 1.2cm},clip]{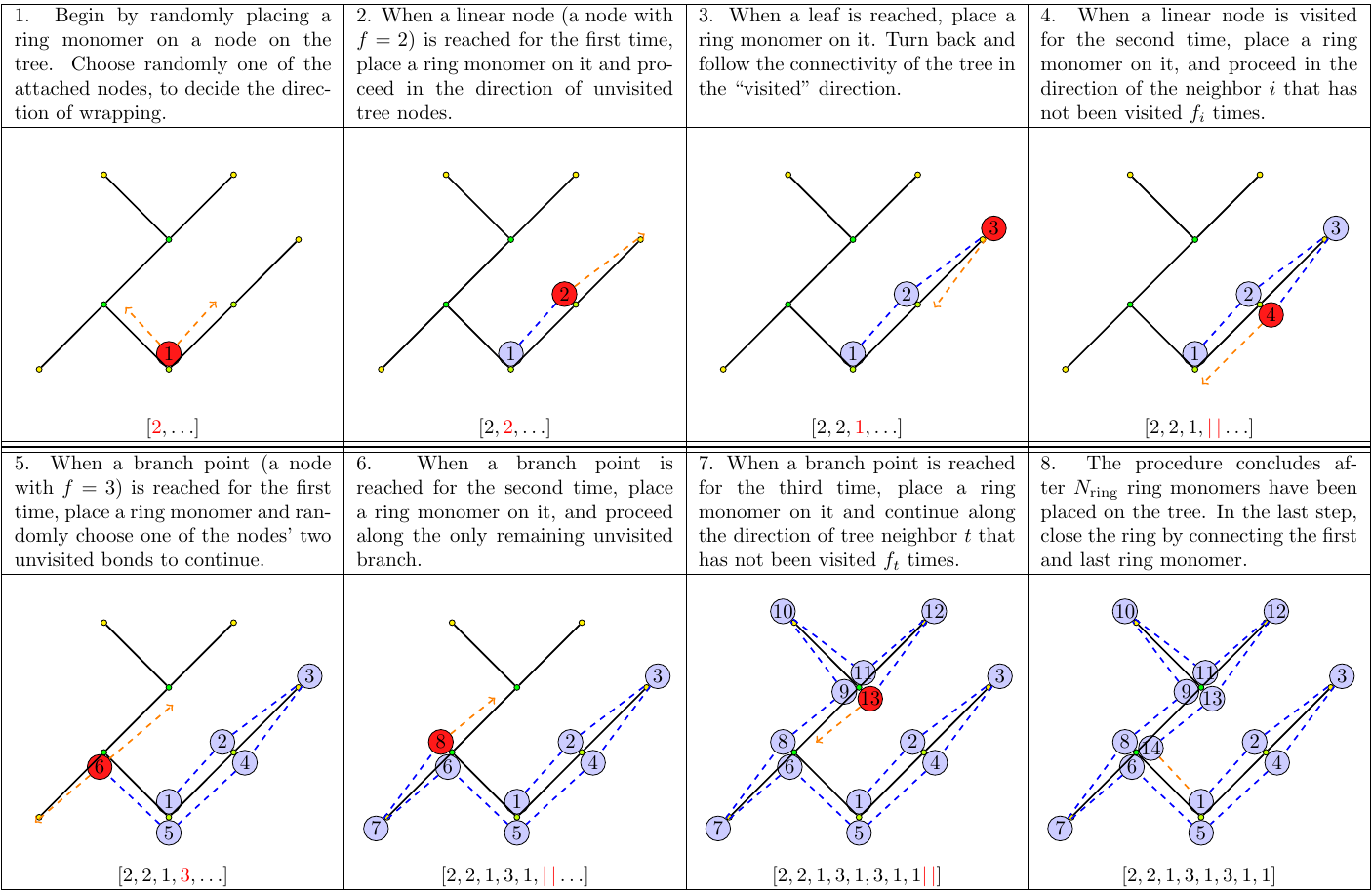}
\caption{
Rules to construct a tightly double-folded ring polymer (violet circles) wrapped around a tree and how to translate this into a corresponding wrapping code.
At each step, the last placed ring monomer is colored in red. 
The possible directions of the wrapping procedure are indicated by the orange arrows. 
In every frame, the evolving construction of the wrapping code is shown in correspondence with the placed ring monomers. 
When a ring monomer is placed that does not correspond to a new entry in the wrapping code, this is indicated by the symbol ``\textcolor{red}{$||$}".
}
\label{fig:Wrapping_procedure}
\end{figure*}
%%%

%%%
{\it Non-interacting trees characterizing double-folding in the absence of volume interactions.}
%%%
In analogy to the standard phantom chain model~\cite{DoiEdwards,RubinsteinColby}, the present work focuses on double-folding in the absence of volume interactions.
While their inclusion in simulations is straightforward~\cite{MadrasJPhysA1992,GrassbergerJPhysA2005,Rosa2016a,Rosa2016b} and their effects can be rationalised by Flory theory~\cite{MaritanGiacometti2013,GrosbergSoftMatter2014,EveraersGrosbergRubinsteinRosa}, exact treatments remain a challenge~\cite{LubenskyIsaacson1979,ParisiSourlasPRL1981,Gujrati1995,JanssenStenullPRE2011}.
Below we distinguish between {\it configurations} and {\it conformations} following the standard notation in the fields of graph theory and polymer physics. In particular, we use the term {\it configuration} to refer to the connectivity of a tree or the secondary structure~\cite{RosaEveraers2019} of a double-folded ring, while we employ {\it conformation} to designate a spatial embedding (Fig.~\ref{fig:Modelfigure}(b)). 
Specifically, we consider double-folding onto {\it acyclic} trees embedded on a common regular lattice of unit step length $b$ and with a coordination number $c$ with $c=2d$ for the hypercubic lattice in $d$ dimensions and $c=12$ for the $3d$ FCC lattice.
While in the general case of ideal random trees the functionality $f$ of tree nodes ({\it i.e.}, the number of other nodes they are connected to) is not restricted, below we mostly focus on the case where $f\le 3$ so that our trees are composed of $\{N_1, N_2, N_3\}$ nodes of functionality $f=1,2,3$, with $N_{\rm tree} = N_1+N_2+N_3$ and the relations
\begin{eqnarray}
N_1(N_{\rm tree},N_3) & = & N_3+2 \, , \label{eq:n1_res} \\
N_2(N_{\rm tree},N_3) & = & N_{\rm tree} - 2N_3 - 2 \, , \label{eq:n2_res}
\end{eqnarray}
between node compositions.
%\footnote{Eqs.~\eqref{eq:n1_res} and~\eqref{eq:n2_res} imply $N_1\geq 2$ and $N_2, N_3 \geq 0$.}
At the end we briefly consider the general case (treated in detail in Section~\ref{sec:arbitrary_functionality} in Supplemental Material (SM~\cite{SMnote})).

%%%
{\it Wrapped trees and a wrapping code}.
%%%
Following the procedure originally outlined in~\cite{RosaEveraers2019} and described in detail in Fig.~\ref{fig:Wrapping_procedure}, double-folding forces the ring to traverse each tree bond twice, so a tree of size $N_{\rm tree}$ is wrapped by a double-folded ring of length 
$
%N_{\rm ring} = 
2 (N_{\rm tree}-1)$.
In the course of the wrapping procedure every tree node $t$ is visited exactly $f_t$ times. %, in agreement with Eq.~\eqref{eq:ring_to_tree}.
Given a tree with $N_3$ branch-nodes, the wrapping can be performed in
\begin{equation}\label{eq:degeneracy of ring labels}
2 (N_{\rm tree}-1) \times 2^{N_3}
\end{equation}
distinct ways, where the factor $2 (N_{\rm tree}-1)$ denotes the number of circular permutations of the ring ({\it i.e.}, the possibility of choosing which ring monomer gets the label $i=1$) and the factor $2^{N_3}$ enumerates the different directions of wrapping that can be chosen when encountering a branch-node ({\it i.e.}, step 5 in Fig.~\ref{fig:Wrapping_procedure}).

The wrapping code records the order in which new tree nodes are encountered along the ring.
Importantly, walking along the labelled ring starting from the ring monomer $i=1$ one can encode the configuration by noting the functionalities (Fig.~\ref{fig:Wrapping_procedure}),
\begin{equation}\label{eq:WrappingCode}
\left [f_1, \dots, f_{N_{\rm tree}} \right ] \, ,
\end{equation}
of newly encountered tree nodes. 
As illustrated in Sec.~\ref{sec:DecodingWrapping} in SM~\cite{SMnote}, a wrapping sequence contains the information to reconstruct the unique configuration of a double-folded ring polymer.
In particular, any permutation in the wrapping sequence corresponds to a different set of monomer pairings or secondary structure.
There is thus a one-to-one correspondence between the configuration of a double-folded ring polymer and the wrapping code.
To specify the polymer's conformation, one must additionally record, during the wrapping process, each newly encountered tree-bond vector, $\{\vec b_1 = \vec r_2-\vec r_1, \ldots, \vec b_{N_{\rm tree}-1}\}$, where $\vec r_i$ is the spatial position of node $i$ (with $i=1, ..., N_{\rm tree}$) in the embedding space.
%

%%%
\begin{figure*}%[ht]
\includegraphics[width=0.75\textwidth]{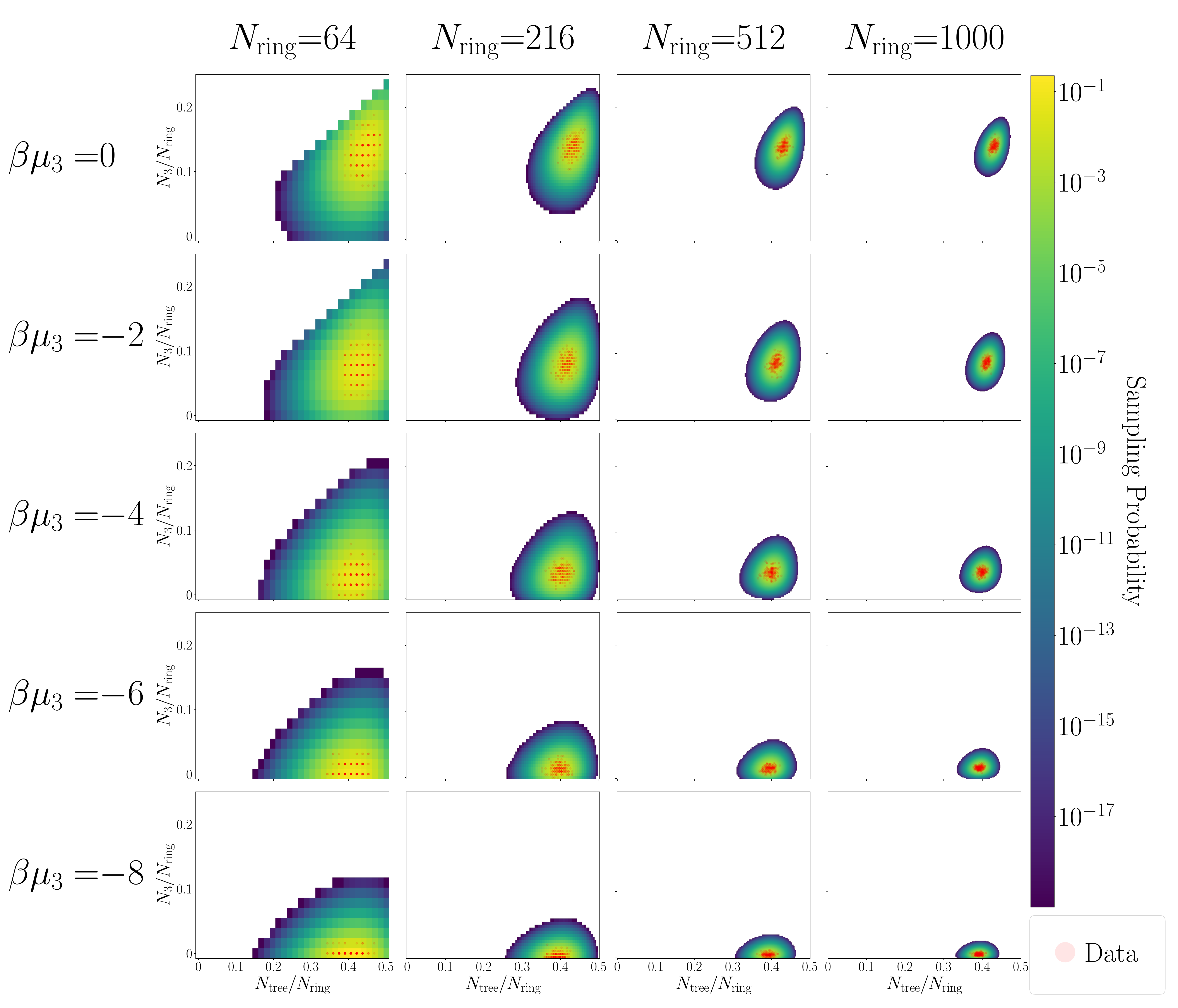}
\caption{
Density map representation of the $2d$ probability distributions (Eq.~\eqref{eq:probability}) of sampling a tuple ($N_{\rm tree}$, $N_3$) at a given $N_{\rm ring}$ and for different values (see legends) of the branching chemical potential $\beta {\treetilde\mu}_3$. %(see Eq.~\eqref{eq:BranchingHam}).
Yellow/blue regions are more/less likely to be sampled, as indicated by the color bar on the right.
Tuples with a probability of sampling $<10^{-19}$ are white in the above plots.
The red dots represent the sampled data points from computer simulations of the elastic polymer model, they all fall on the maxima of the predicted probability distributions with no systematic deviations observed. 
}
\label{fig:probability_distributions}
\end{figure*}
%%%

%%%
{\it The total number of viable wrapping codes.}
%%%
Not every generic $N_{\rm tree}$-long succession of $1$'s, $2$'s and $3$'s corresponds to a legitimate double-folded ring.
For instance, the procedure illustrated in Fig.~\ref{fig:Wrapping_procedure} implies that a proper wrapping code must always end with $f_{N_{\rm tree}} = 1$, because the tree node with index $=N_{\rm tree}$ is by definition only visited for the first (and hence only) time towards the end of the wrapping procedure. 
For given values $(N_{\rm tree}, N_3)$, this constraint leaves
\begin{equation}\label{eq:counting_first}
\frac{(N_{\rm tree}-1)!}{(N_1-1)! \, N_2! \, N_3!} 
%= \frac{(N_{\rm tree}-1)!}{(N_3+1)! \, (N_{\rm tree} - 2N_3 - 2)! \, N_3!}
\end{equation}
potential orderings of the first $N_{\rm tree}-1$ entries of a wrapping code.

The condition that the ring must only be closed with the last placed bond rules out a code like $[1,1,3,1]$, where the ring would already be closed after two steps. 
In particular, for a viable wrapping code the number of 3's may never be ahead of the number of 1's, when the first $N_{\rm tree}-1$ entries are read backwards.
The probability for this to be the case in a random sequence of $N_1-1=N_3+1$ 1's and $N_3$ 3's is given by a generalized form of Bertrand's ``ballot theorem''~\cite{Bertrand1887} that permits {\it ties} (see Sec.~\ref{Sec:Ballot_Theorem} in SM~\cite{SMnote} for details), namely the fraction of valid sequences is
\begin{equation}\label{eq:Ballot_applied}
\frac2{N_3+2} \, .
\end{equation}
Multiplying the total number of admissible permutations in Eq.~\eqref{eq:counting_first} by the ballot-theorem fraction in Eq.~\eqref{eq:Ballot_applied}, we obtain the total number of viable wrapping codes and thus of tightly double-folded rings configurations for given $(N_{\rm tree}, N_3)$,
\begin{equation}\label{eq:labeled_tree_wrappings_labeled_ring}
\Omega_{\rm ring}(N_{\rm tree}, N_3) =\frac{2(N_{\rm tree}-1)!}{N_1! \, N_2! \, N_3!} \, ,
%= \frac{2 \, (N_{\rm tree} -1)!}{(N_3+2)! \, (N_{\rm tree}-2N_3-2)! \, N_3!} \, .
\end{equation}
with $N_1$ and $N_2$ given by Eqs.~(\ref{eq:n1_res}) and (\ref{eq:n2_res}).

%%%
%\subsection{Controlling the branching of double-folded rings}\label{sec:mu_branching}
{\it Controlling the branching activity.}
%%%
The branching activity of the double-folded rings can be controlled via a chemical potential ${\treetilde\mu}_3$~\cite{Rosa2016a,Rosa2016b,Rosa2016c,Amoebapaper2024}
in the partition function for tightly double-folded rings,
\begin{eqnarray}\label{eq:Z_ring}
Z_{\rm ring}
& = & Z_{\rm ring}(N_{\rm tree}, {\treetilde\mu}_3) \\
& = & c^{N_{\rm tree}-1} \sum_{N_3=0}^{N_{3, {\rm max}}} \Omega_{\rm ring}(N_{\rm tree},N_3) \exp{(\beta {\treetilde\mu}_3 N_3)} \, , \nonumber
\end{eqnarray}
where $N_{3, {\rm max}}=(N_{\rm tree}-2)/2$ for $N_{\rm tree}$ even and $N_{3, {\rm max}}=(N_{\rm tree}-3)/2$ for $N_{\rm tree}$ odd (see Eq.~\eqref{eq:n2_res}) and $c$ is the coordination number of the embedding lattice.

%%%
{\it Accounting for the reptons in the elastic lattice model.}
%%%
The elasticity in our lattice model~\cite{Ghobadpour2021,Ghobadpour2025} of tightly double-folded rings emerges from the ring-bond length being either zero or equal to the lattice spacing (see Fig.~\ref{fig:Modelfigure}(a3)).
However, up to this point, the theory was formulated at fixed tree size $N_{\rm tree}$.
To compare with the elastic lattice model, we now switch to an ensemble at fixed ring length $N_{\rm ring}$ in which the ring-bond length, and so the size $N_{\rm tree}$ of the underlying tree, fluctuates.
Now, the probability to observe a double-folded ring conformation with $N_{\rm tree}$ and $N_3$ is given by
\begin{align}\label{eq:probability}
p(N_{\rm tree}, N_3 | &N_{\rm ring}, {\treetilde\mu}_3) = \\
& \frac{ \Omega_{\rm rep}(N_{\rm ring}, N_{\rm tree}) \, \Omega_{\rm ring} (N_{\rm tree} , N_3) \, e^{\beta {\treetilde\mu}_3 N_3}}{Z_{\rm elastic}(N_{\rm ring}, {\treetilde\mu}_3)} \, , \nonumber
\end{align}
where $\Omega_{\rm rep}(N_{\rm ring}, N_{\rm tree})$ is the number of ways to place zero-length bonds (`reptons') on a ring (see Sec.~\ref{sec:ReptonInclusion} in SM~\cite{SMnote}) and where
\begin{equation}\label{eq:ElasticModel-Z}
Z_{\rm elastic}(N_{\rm ring}, {\treetilde\mu}_3) = \sum_{N_{\rm tree}} \Omega_{\rm rep}(N_{\rm ring}, N_{\rm tree}) \, Z_{\rm ring}(N_{\rm tree}, {\treetilde\mu}_3)
\end{equation}
is the partition function of the elastic system.
Expectation values for the tree size and the number of $f=1,2,3$-functional nodes as a function of $(N_{\rm ring}, {\treetilde\mu}_3)$ can be calculated by numerically summing 
\begin{eqnarray}
\langle N_{\rm tree} \rangle & = & \sum_{N_{\rm tree}} \sum_{N_3}  N_{\rm tree} \, p(N_{\rm tree}, N_3 | N_{\rm ring}, {\treetilde\mu}_3) \, , \label{eq:ElasticModel-Ntree} \\
\langle N_{3} \rangle & = & \sum_{N_{\rm tree}}  \sum_{N_3} N_{3}  \,   p(N_{\rm tree}, N_3 | N_{\rm ring}, {\treetilde\mu}_3) \, . \label{eq:ElasticModel-N3}
\end{eqnarray}
The values for $\langle N_{1} \rangle$ and $\langle N_{2} \rangle$ follow then from Eqs.~\eqref{eq:n1_res} and~\eqref{eq:n2_res}.
For more details see Sec.~\ref{sec:ReptonInclusion} in SM~\cite{SMnote}.

%%%
{\it Comparison to simulation results.}
%%%
To validate our results we have performed computer simulations of our elastic lattice model~\cite{Ghobadpour2021,Ghobadpour2025} for tightly double-folded rings, where we included the possibility into the code to control the branching activity along the lines of Eq.~\eqref{eq:Z_ring}.
We have generated double-folded rings polymers for lengths $N_{\rm ring} \in [64,216,512,1000]$ and chemical potentials $\beta {\treetilde\mu}_3 \in [0,-2,-4,-6,-8]$.
All simulated samples consist of $200$ independent double-folded ring conformations.
For details of the implementation, the equilibration procedure and the simulations we refer the reader to Refs.~\cite{Ghobadpour2021,Ghobadpour2025}.

%%%
\begin{figure}%[ht]
\includegraphics[width=0.48\textwidth]{lambda_plots.pdf}
\caption{
Expectation values of node functionalities normalized to the mean tree size $\langle N_{\rm tree}\rangle$ (main panel) and mean tree size normalized to the ring size $N_{\rm ring}$ (inset) as a function of the branching chemical potential $\beta\mu_3$.
Results:
($\times$) sampled averages of the elastic lattice model;
($\square$) numerically evaluated sums Eqs.~\eqref{eq:ElasticModel-Ntree} and~\eqref{eq:ElasticModel-N3};
(lines) analytical asymptotic formulas Eqs.~\eqref{eq:AsymptoticNtree}-\eqref{eq:AsymptoticN2}.
}
\label{fig:lambdaplots}
\end{figure}
%%%

As visually represented in Fig.~\ref{fig:probability_distributions}, the sampled datapoints (red dots) all fall around the peaks of the theoretical distributions, Eq.~\eqref{eq:probability}, into bins with expected sampling probabilities $p(N_{\rm tree}, N_3 | N_{\rm ring}, {\treetilde\mu}_3) \gtrsim 10^{-2}$.
A more rigorous validation, based on a statistical quantitative test, that Eq.~\eqref{eq:probability} correctly represents the underlying distribution of sampled datapoints is described in detail in Sec.~\ref{sec:test_statistics} in SM~\cite{SMnote}.
Finally, Fig.~\ref{fig:lambdaplots} shows (main panel) the corresponding averages for $\langle N_1 + N_3\rangle /2$ and $\langle N_{2} \rangle$ normalized to $\langle N_{\rm tree} \rangle$ as well as (inset) for $\langle N_{\rm tree} \rangle - 1$ normalized to $N_{\rm ring}/2$. 
Not surprisingly, the sampled averages ($\times$) are also in good agreement with the numerically evaluated sums ($\square$).

%%%
{\it Analysis of the asymptotic behavior.}
%%%
In Sec.~\ref{sec:AsymptoticsZring} in SM~\cite{SMnote} we show that the partition function for double-folded rings, Eq.~\eqref{eq:Z_ring}, is asymptotically dominated by the term
\begin{equation}\label{eq:AsymptoticZring}
Z_{\rm ring}(N_{\rm tree}, {\treetilde\mu}_3) \sim \left( \frac{c}{1-2\lambda} \right)^{N_{\rm tree}} \, ,
\end{equation}
with
\begin{equation}\label{eq:lambda}
\lambda({\treetilde\mu}_3) = (2+\exp( -\beta {\treetilde\mu}_3 / 2))^{-1} \ .
\end{equation}
Using Stirling's approximation for $\Omega_{\rm rep}(N_{\rm ring}, N_{\rm tree})$ (Eq.~\eqref{eq:ElasticModel-ChoosingBonds} in SM~\cite{SMnote}) and keeping only the dominant tree size in Eq.~(\ref{eq:ElasticModel-Z}) one finds (see Sec.~\ref{sec:expectedtreesize} in SM~\cite{SMnote}) 
\begin{equation}\label{eq:AsymptoticNtree}
%\left \langle N_{\rm tree}\right \rangle \approx 
%1 + \frac{\left \langle N_{\rm rep}\right \rangle}{2}\sqrt{ \left(1+2\exp\left(\frac{\beta {\treetilde\mu}_3}{2}\right)\right) c} 
\frac{ \left \langle N_{\rm tree}\right \rangle -1}{N_{\rm ring}/2} \approx \left( \sqrt{\frac{1-2\lambda}c} +1 \right)^{-1}
\end{equation}
and Eqs.~(\ref{eq:ElasticModel-Ntree}) and (\ref{eq:ElasticModel-N3}) can be approximated as
\begin{equation}\label{eq:AsymptoticN1N3}
\frac{ \left\langle N_1 \right\rangle -1 } {\left\langle N_{\rm tree}\right\rangle} = \frac{ \left\langle N_3 \right\rangle +1 } {\left\langle N_{\rm tree}\right\rangle} \approx \lambda \, ,
\end{equation}
so that
\begin{equation}\label{eq:AsymptoticN2}
\frac{ \left\langle N_2 \right\rangle } {\left\langle N_{\rm tree}\right\rangle} \approx 1-2\lambda \, . 
\end{equation}
The corresponding lines in Fig.~\ref{fig:lambdaplots} (main panel and inset) are in excellent agreement with the numerical data points for the explored values of $N_{\rm ring}$.
Similarly, we have also found excellent agreement between the predicted distribution of reptons on nodes of given functionality (which is described by a beta-binomial law, see Eq.~\eqref{eq:repton_distribution} in SM~\cite{SMnote}) and the numerical data (for an in-depth discussion, see Sec.~\ref{sec:ReptonDistrPerNode} in SM~\cite{SMnote}).

%%%
\begin{figure}%[ht]
\includegraphics[width=0.48\textwidth]{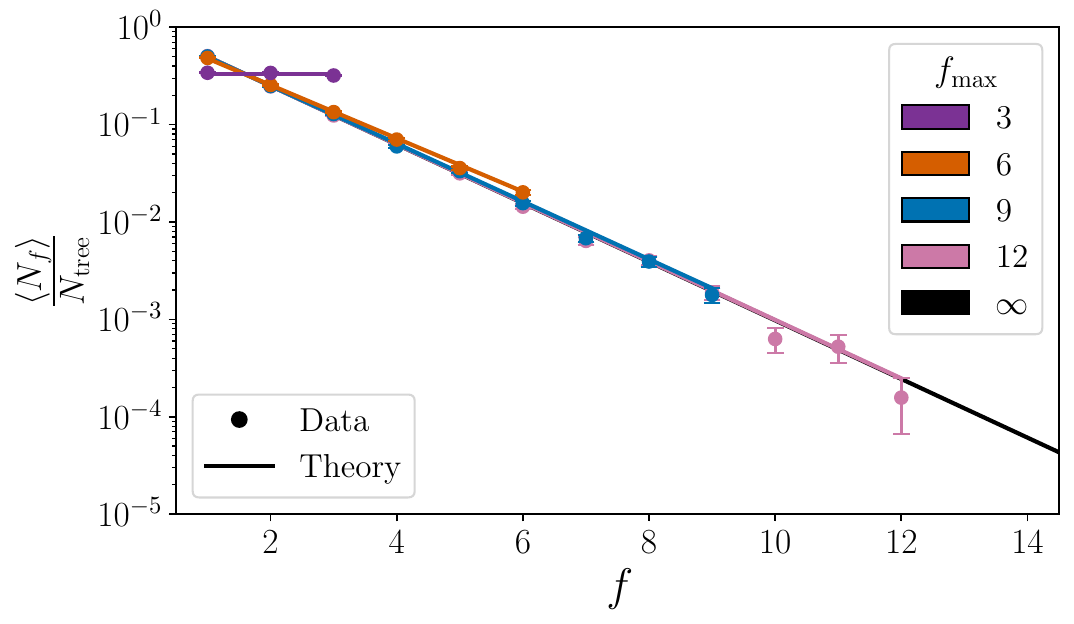}
\caption{
Mean number of nodes of functionality $f$ for trees with maximal nodes' functionality $f_{\rm max}$ and branching chemical potentials $\{ {\treetilde\mu}_f = 0 \}$.
Symbols and lines are for sampled averages of the elastic lattice model and our theory.
}
\label{fig:functionalityplot}
\end{figure}
%%%

%%%%
{\it Arbitrary node functionalities.}
%%%%
In general, a ring wrapping a tree passes each tree node $t$ exactly $f_{t}$ times.
This implies that the wrapping procedure has $(f_{t}-1)!$ choices of direction for wrapping per tree node $t$.
Reversing the arguments for Eq.~\eqref{eq:labeled_tree_wrappings_labeled_ring} and Sec.~\ref{sec:DecodingWrapping} in SM~\cite{SMnote}, by assigning to each labeled tree a multiplicity of $(f_t-1)!$ per tree node of functionality $f_t$, we {\it conjecture} that the generalized version of Eq.~\eqref{eq:labeled_tree_wrappings_labeled_ring} is given by:
\begin{equation}\label{eq:Multiplicity_with_f>3}
\Omega_{\rm ring}(N_{\rm tree},\{N_f\}) = 2\frac{(N_{\rm tree}-1)!}{N_1!\dots N_{f_{\rm max}}!} \, ,
\end{equation}
where $f_{\rm max}$ is nodes' maximal functionality.
From Eq.~\eqref{eq:Multiplicity_with_f>3}, all previously derived results can be extended accordingly.
In Sec.~\ref{sec:arbitrary_functionality} in SM~\cite{SMnote}, we derive a polynomial identity, Eq.~\eqref{eq:alphapolynomial_gen} in SM~\cite{SMnote}, that needs to be solved numerically to approximate the expected number of tree-nodes $\left\langle N_f \right \rangle$, for given functionality $f$ and corresponding set of branching chemical potentials $\{ {\treetilde\mu}_1, \dots, {\treetilde\mu}_{f_{\rm max}} \}$. 
Notably, there are two simple limiting cases for chemical potentials $\{ {\treetilde\mu}_f = 0 \}$:
\begin{equation}\label{eq:muf>3-LimitingCases}
\lim_{N_{\rm tree}\rightarrow\infty}
\frac{\left\langle N_f \right \rangle}{N_{\rm tree}} =
\left\{
\begin{array}{cc}
1/3 & \mbox{ if $f_{\rm max} = 3$ } \\
\\
2^{-f} & \mbox{ if $f_{\rm max}=\infty$ } \\
\end{array}
\right.
\end{equation}
In Fig.~\ref{fig:functionalityplot} we show that data from the Supporting Information of Ref.~\cite{Ghobadpour2025} for $N_{\rm ring}=216$, $\{ {\treetilde\mu}_f = 0 \}$ and maximal functionalities up to $f_{\rm max} =12$ (symbols) are in good agreement with the theoretical expressions (lines).
The (essentially) equal prevalence of $f=1,2,3$ functional nodes for $f_{\rm max}=3$ corresponds to the crossing of the lines in the main panel of Fig.~\ref{fig:lambdaplots},
while the functionality distributions for $ f_{\rm max} >3$ quickly approach the asymptotic $2^{-f}$ behavior.

%%%
%\section{Conclusion and Outlook}\label{sec:Conclusion}
{\it Conclusion and outlook.}
%%%
To summarize, we have introduced a wrapping code for characterising the secondary structure of double-folded rings, which allowed us to count the number of available configurations in the absence of volume interactions.
Resulting predictions for ensembles with controlled branching activity are in excellent agreement with data from Monte Carlo simulations of an elastic lattice model of non-interacting tightly double-folded rings.
In a forthcoming publication~\cite{CoherentModeling} we will establish a coherent framework for the modelling double-folded rings on the ring and on the tree level, which is based on a direct mapping between the two ensembles explored here and in Ref.~\cite{vanderHoek2025}. 
This connection enables a major gain in computational efficiency by leveraging earlier designed efficient algorithms for randomly branched polymers~\cite{Amoebapaper2024}, unlocking the possibility to explore the consequences of topological constraints on genome organization~\cite{EGPhDThesis}.
%\RalfComment{Forward cite Gabin's paper?}
%\AngeloComment{Maybe, I would not spoil it!}

%%%
{\it Supplemental Material.}
%%%
The supplemental material contains:
a section illustrating the decoding of the wrapping code,
a presentation with demonstration of the ``ballot theorem'',
the derivation of the statistical weight of reptons,
a validation of Eq.~\eqref{eq:probability} using a statistical test,
formulas for the trees/rings asymptotic behavior,
treatment of trees with nodes of arbitrary functionality,
additional figures.

%%%
%\section*{Acknowledgements}
%%%
{\it Acknowledgements.}
PHWvdH acknowledges financial support from PNRR\_M4C2I4.1.\_DM351 funded by NextGenerationEU and the kind hospitality of the ENS-Lyon.
AR acknowledges financial support from PNRR Grant CN\_00000013\_CN-HPC, M4C2I1.4, spoke 7, funded by Next Generation EU.
%%%

%%%
{\it Conflict of interest.}
%%%
The authors have no conflicts to disclose.

%%%
{\it Data availability.}
%%%
The data that support the findings of this study are available from the corresponding author upon reasonable request.

%%%
\bibliography{../biblio.bib}% Produces the bibliography via BibTeX.
%%%

%%%
%\begin{appendix}
%%%

%%%
\clearpage
%%%

%%%
\widetext
\clearpage
\begin{center}
\textbf{\Large Supplemental Material \\ \vspace*{1.5mm} Configurational entropy of randomly double-folding ring polymers} \\
\vspace*{5mm}
Pieter H. W. van der Hoek, Angelo Rosa, Elham Ghobadpour, Ralf Everaers
\vspace*{10mm}
\end{center}
%\balancecolsandclearpage
%%%

%%%%%%%%% Prefix a "S" to all equations, figures, tables and reset the counter %%%%%%%%%%
\setcounter{equation}{0}
\setcounter{figure}{0}
\setcounter{table}{0}
\setcounter{page}{1}
\setcounter{section}{0}
\setcounter{page}{1}
\makeatletter
\renewcommand{\theequation}{S\arabic{equation}}
\renewcommand{\thefigure}{S\arabic{figure}}
\renewcommand{\thetable}{S\arabic{table}}
\renewcommand{\thesection}{S\arabic{section}}
%\renewcommand{\thepage}{S\arabic{page}}
%%%

\tableofcontents
%\addcontentsline{toc}{chapter}{Appendices}
%\addtocontents{toc}{\protect\setcounter{tocdepth}{1}}
%\input{myTOC.toc}

%%%
\clearpage
%%%

%%%
\begin{figure}[ht]
\centering
\includegraphics[page=1,width=0.75\textwidth,trim={3.2cm 2.2cm 2.7cm 1.2cm},clip]{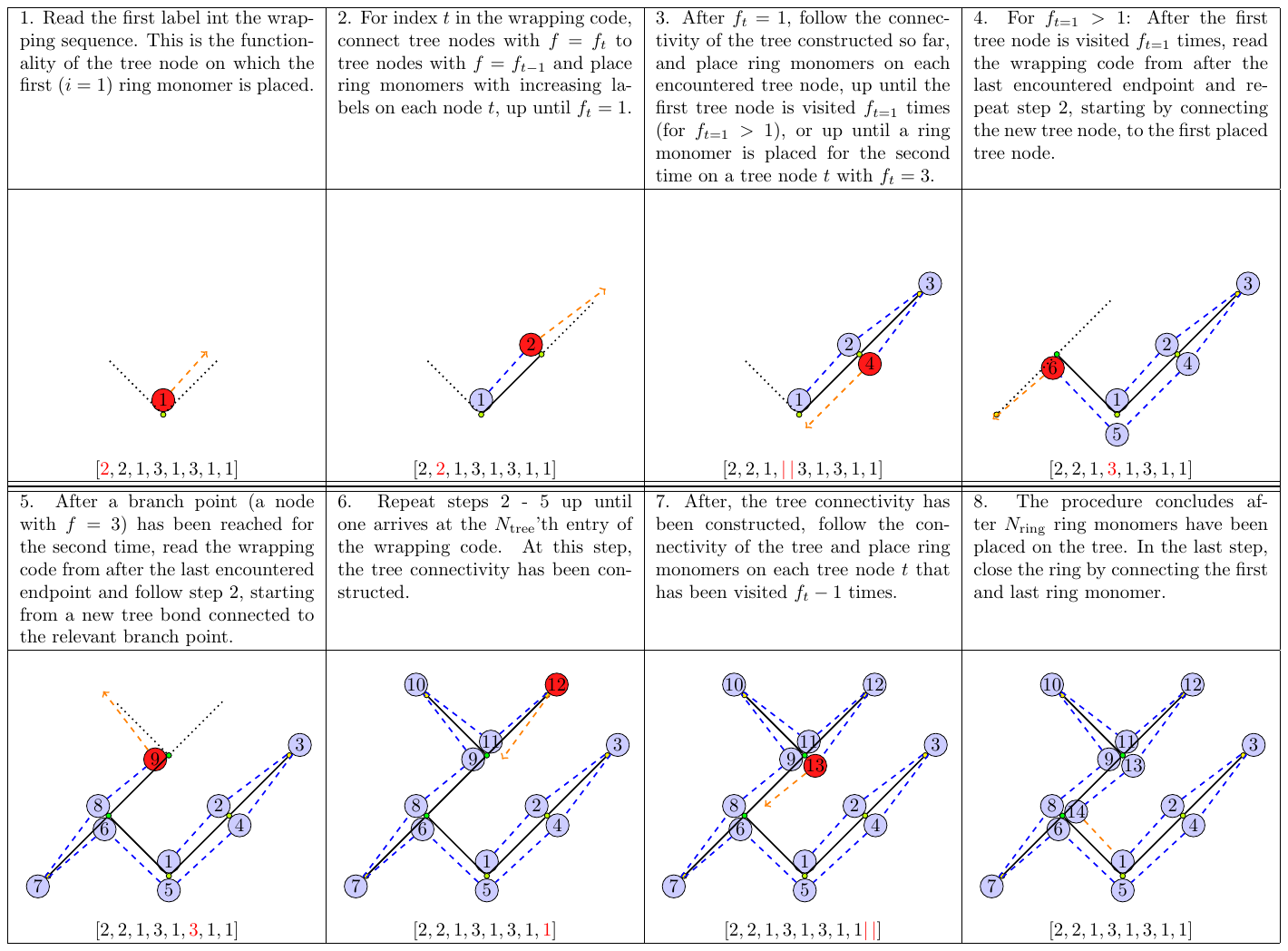}
\caption{
Reverse procedure to construct a tightly double-folded ring polymer from a given wrapping code.
Colorcode is as in Fig.~\ref{fig:Wrapping_procedure} of main text.
}
\label{fig:Decoding}
\end{figure}
%%%

%%%%
\section{Decoding the wrapping code}\label{sec:DecodingWrapping}
%%%% 
The reverse of the wrapping procedure described in main text, namely the {\it decoding} of a wrapping code, is displayed in detail in Fig.~\ref{fig:Decoding}.
After placing the first ring monomer at spatial position $\vec r_1$, the ring and the tree are extended in the direction of the listed bond vectors by placing monomers and tree nodes of the functionality indicated in the code.
After placing a $1$-functional node, only the ring is extended retracing the already placed nodes to satisfy the double-folding constraint.
These two operations are iterated for the remaining code.

%%%
\section{Bertrand's ``ballot theorem''}\label{Sec:Ballot_Theorem}
%%%
The Bertrand's ``ballot theorem'' bears the name of Joseph Louis Fran{\c{c}}ois Bertrand who originally posed the following problem in the context of elections and counting voting ballots~\cite{Bertrand1887}:
\\

{\it
Consider an election between two candidates $A$ and $B$, where $V_A$ and $V_B$ are, respectively, the number of ballots in favor of candidate $A$ and candidate $B$.
If $V_A > V_B$ and one counts the ballots one after the other, in how many scenarios is candidate $A$ always ahead of $B$ during the counting process?
}
\\

Bertrand demonstrated that this number is given by the fraction
\begin{equation}\label{eq:Ballot_original}
\frac{V_A - V_B}{V_A + V_B}
\end{equation}
of all possible counting processes. 

%%%
%\clearpage
%%%
\begin{figure}[ht] 

\begin{tikzpicture}[scale=0.9]
\begin{axis}[
    axis lines =center,
    xlabel={$t$},
    ylabel={$\Delta_{AB}$},
    xmin=0, xmax=10.5,
    ymin=-3, ymax=6.5,
    xtick={0,1,2,3,4,5,6,7,8,9,10},
    ytick={-3,-2,-1,0,1,2,3,4,5,6}  
]

\addplot[
    color=green,
    mark=square,
    ]
    coordinates {
    (0,0)(1,1)(2,2)(3,3)(4,4)(5,3)(6,2)(7,1)(8,2)(9,3)(10,2)
    };
\addplot[
    color=red,
    mark=square,
    ]
    coordinates {
    (0,-0.01)(1,0.99)(2,1.99)(3,0.99)(4,-0.01)(5,-1.01)(6,-2.01)(7,-1.01)(8,-0.01)(9,0.99)(10,1.99)
    };]
\addplot[dashed,
    color=violet,
    mark=square,
    ]
    coordinates {
    (0,0.01)(1,-0.99)(2,-1.99)(3,-0.99)(4,0.01)(5,-0.99)(6,-1.99)(7,-0.99)(8,0.01)(9,0.99)(10,1.99)
    };]
\addplot[dotted,
    color=black,
    ]
    coordinates {
    (0,1)(10,1)
    };]
    
\node (source) at (axis cs:9,5){};
       \node (destination) at (axis cs:10,6.3){$A$};
       \draw[->](source)--(destination);
\node[minimum height=0.1cm,draw](source) at (axis cs:9,5){};
       \node (destination) at (axis cs:10,3.7){$B$};
       \draw[->](source)--(destination);
    
\end{axis}

\end{tikzpicture}
\caption{
Random-walk representation of ballots counting for $V_A = 6$ and $V_B = 4$.
The green path shows $A$ always ahead.
The red path ties at $t = 4$.
The dashed violet path is the reflection of the red path over the $t$-axis up to $t = 4$.
}
\label{fig:Random_Walk}
\end{figure}
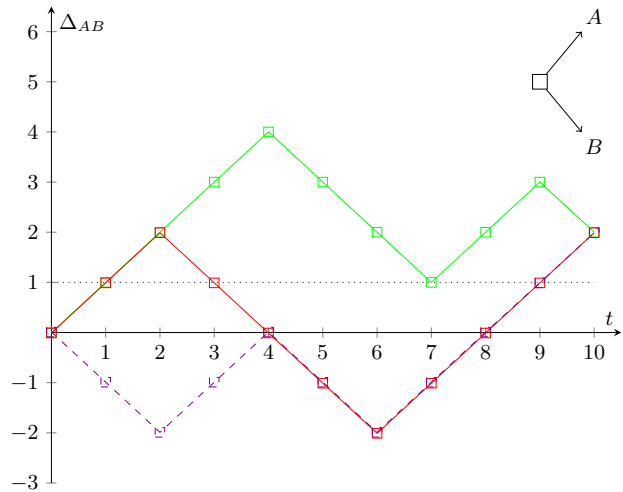
%%%

{\it Proof of the theorem:}
The ``ballot theorem'' has been proven several times since the late 1800s~\cite{Bertrand1887,Barbier1887,Andre1887,Appli1923,Renault2007}.
For its simplicity, here we follow the approach along the lines of Ref.~\cite{Andre1887} and visualize the counting of ballots as a $1d$ random walk over time~\cite{Appli1923}, where the $x$-axis represents time $t$ ({\it i.e.}, the number of ballots counted), and the $y$-axis tracks the vote difference $\Delta_{AB}(t) = V_A(t) - V_B(t)$ (see Fig.~\ref{fig:Random_Walk}):
\begin{itemize}
\item
A vote for $A$ corresponds to a step up.
\item
A vote for $B$ corresponds to a step down.
\end{itemize}
All paths start at $(0,0)$ and end at $(V_A + V_B, V_A - V_B)$.
A valid ``always-ahead'' path corresponds to a walk that:
\begin{itemize}
\item
Starts with a step up.
\item
Never drops below the line $\Delta_{AB} = 1$ (the horizontal dotted line in Fig.~\ref{fig:Random_Walk}) after the first step.
\end{itemize}
Consider now the ensemble of random walks that start with a move up, whose fraction is given by
\begin{equation}\label{eq:V_A_fraction}
\frac{V_A}{V_A +V_B} \, .
\end{equation}
Any such walk that eventually intersects the $t$-axis at point $t=p$ can be reflected (dashed violet path in Fig.~\ref{fig:Random_Walk}) around the $t$-axis between points $(0,0)$ to $(p,0)$ to yield a walk that starts with a down step. 
Since $V_A > V_B$, any random walk that starts with a step down intersects the $t$-axis at a later point $(p,0)$, with $p>0$.
There is thus a one-to-one correspondence between the ensemble of walks starting with a step up that eventually intersect the $t$-axis and that of walks starting with a step down.
Since the fraction of ``start-down'' walks is %``start-up" walks that ever hit the $t$-axis is
\begin{equation}\label{eq:V_B_fraction}
\frac{V_B}{V_A + V_B} \, ,
\end{equation}
$1$ minus twice this fraction gives back Eq.~\eqref{eq:Ballot_original}, thus proving the theorem.

{\it Ties allowed:}
The ``ballot theorem'' can be generalized for a counting of votes with ties allowed, {\it i.e.} paths that never go below $\Delta_{AB} = 0$.
In fact, if we fix the first move to be up in the previously discussed scenario, the remaining $V_A + V_B - 1$ steps correspond to a ballot problem between $V_A' = V_A - 1$ votes for candidate $A$ and $V_B' = V_B$ votes for candidate $B$, with ties allowed on the ``$\Delta_{AB} = 1$''-line.
Therefore, for $V_A' \geq V_B'$ the fraction of scenarios with ties allowed is given by Eq.~\eqref{eq:Ballot_original} divided by Eq.~\eqref{eq:V_A_fraction},
\begin{equation}\label{eq:Ballot_ties}
\frac{(V_A - V_B) / (V_A + V_B) }{V_A / (V_A + V_B)} = \frac{V_A-V_B}{V_A} = \frac{V_A' - V_B' + 1}{V_A'+1} \, .
\end{equation}

{\it Including blank votes:}
A simple generalization of the process with ties consists in the introduction of $V_C'$ ``blank votes'' which, quite obviously, do not represent any vertical movement in the analogous random walk. 
Since these horizontal steps can be inserted at any point in the random walk, it is still possible to use the same reflection argument that we have exploited before. 
For these reasons, Eq.~\eqref{eq:Ballot_ties} remains valid even if $V_C'$ blank votes are present.
The connection to the total number of viable wrapping codes (see main text) %Sec.~\ref{sec:NumberViableWrappingCodes})
should be now clear: as the the number of 3's may never be ahead of the number of 1's, this is equivalent to set $V_A' = N_3 + 1$ and $V_B' = N_3$ in Eq.~\eqref{eq:Ballot_ties} (with $V_C' = N - 2N_3 - 2$), thus proving Eq.~\eqref{eq:Ballot_applied} of main text.

%%%
\section{Introducing reptons}\label{sec:ReptonInclusion}
%%%
The elastic model \cite{Ghobadpour2021,Ghobadpour2025} contains units of stored length (see Fig.~\ref{fig:Modelfigure}(a3) of main text), which are missing in Eq.~\eqref{eq:Z_ring} of main text.
To compare the results of the simulations to theory, Eq.~\eqref{eq:Z_ring} of main text must be then modified to include the presence of such units.
In this context, the tree structure of $N_{\rm tree}$ tree nodes is ``complemented'' by $N_{\rm rep}$ reptons hosting each a unit of stored length.
Following that a tree of size  $N_{\rm tree}$ can be wrapped without reptons by $2(N_{\rm tree} -1)$ ring monomers (see main text), %(see Sec.~\ref{sec:WrappedTrees}),
we have the constraint:
\begin{equation}\label{eq:repton_constraint}
N_{\rm rep} = N_{\rm ring}-2(N_{\rm tree}-1) \, .
\end{equation}
Assuming the distribution of reptons is random along the ring, there are therefore
\begin{equation}\label{eq:ElasticModel-ChoosingBonds}
\Omega_{\rm rep}(N_{\rm ring},N_{\rm tree}) = \frac{N_{\rm ring}!}{(2(N_{\rm tree}-1))! \,  (N_{\rm rep})!} = \frac{N_{\rm ring}!}{(2(N_{\rm tree}-1))! \,  (N_{\rm ring}-2(N_{\rm tree}-1))!}
\end{equation}
distinct possible ways of placing reptons hosting each a unit of stored length.
%Finally, the total multiplicity of double-folded rings around trees is given by (see Eq.~\eqref{eq:labeled_tree_wrappings_labeled_ring}):
%
%\begin{widetext}
%\begin{equation}\label{eq:ElasticModel-Multiplicity}
%\Omega_{\rm ring}(N_{\rm ring}, N_{\rm tree}, N_3) =
%\left\{
%\begin{array}{ll}
%1 \, , & \text{for } N_{\rm tree} = 1 \\
%\\
%c^{N_{\rm tree}-1} \, \frac{N_{\rm ring}!}{(2(N_{\rm tree}-1))! \, N_{\rm rep}!} \, \frac{2 \, (N_{\rm tree} -1)!}{(N_3+2)! \, (N_{\rm tree}-2N_3-2)! \, N_3!} \, , & \text{for } N_{\rm tree} > 1
%\end{array}
%\right.
%\end{equation}
%\end{widetext}
%
%where $c=12$ for the $3d$ FCC lattice.

%%%
\section{Statistical test for validation of distribution functions $p(N_{\rm tree}, N_3 | N_{\rm ring}, {\treetilde\mu}_3)$}\label{sec:test_statistics}
%%%
We remind the reader that, for chosen values $(N_{\rm ring}, {\treetilde\mu}_3)$ ({\it i.e.}, for each of the $20$ panels of Fig.~\ref{fig:probability_distributions} of main text), the simulated sample consist of $200$ independent double-folded ring conformations ({\it i.e.}, the red dots in each panel of Fig.~\ref{fig:probability_distributions} of main text).
Then, a more stringent statistical test that confirms that the analytical distribution $p(N_{\rm tree}, N_3 | N_{\rm ring}, {\treetilde\mu}_3)$ (Eq.~\eqref{eq:probability} of main text) is compatible with the observed simulated sample is given as the following. %, we report $p_{\rm logL}$-values. %These values quantify how likely the observed data are under the theoretical probability distribution. %The procedure is as follows.

%%%
\begin{figure}%[ht]
\includegraphics[width=0.95\textwidth]{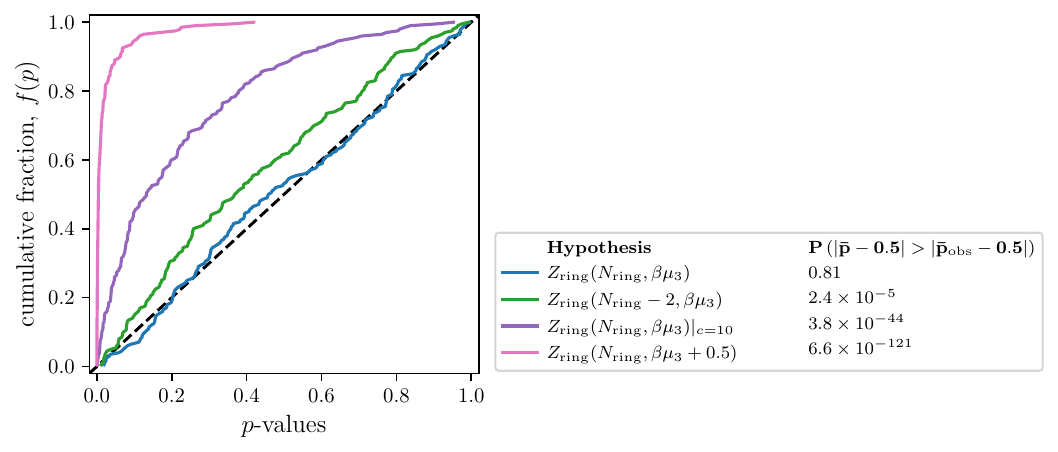}
\caption{
(Left)
Cumulative fraction of $p$-values smaller than a threshold value $\in[0,1]$.
Only for the correct choice of the model parameters (blue line) the fraction lies along the ``$y=x$''-line.
Conversely, even small deviations from the correct parameter values ({\it e.g.}, $N_{\rm ring}$, $\mu_3$, or the lattice coordination number $c$; see legend) lead to significant departures from the ``$y=x$'' behavior.
(Right)
Accordingly, the probability of observing an average $p$-value ($\bar p$) as far away from $0.5$ as the observed one (${\bar p}_{\rm obs}$) becomes rapidly negligible.
}
\label{fig:pNtreeN3-Validation}
\end{figure}
%%%

%
\begin{enumerate}
\item
Arrange the entire ``$200 \times 20$'' observed sample in $200$ distinct arrays of the form:
\begin{equation}\label{eq:DefineDataSetForLogLikelihood}
\left\{ \left[ N_{\rm tree}\left(N_{{\rm ring}}^{(i)}, {\treetilde\mu}_3^{(i)}\right), N_3\left(N_{{\rm ring}}^{(i)}, {\treetilde\mu}_3^{(i)}\right) \right]_{i=1,20} \right\} \, ,
\end{equation}
where each entry in~\eqref{eq:DefineDataSetForLogLikelihood} consists of one $(N_{\rm tree}, N_3)$ pair for every $(N_{\rm ring}, {\treetilde\mu}_3)$.
\item
Evaluate the so-called {\it log-likelihood} of each array, defined as:
\begin{equation}\label{eq:logL}
\mbox{log-likelihood} \equiv \sum_{i=1}^{20} \log\!\left[ \, p \!\left( N_{\rm tree}\left(N_{{\rm ring}}^{(i)}, {\treetilde\mu}_3^{(i)}\right), N_3\left(N_{{\rm ring}}^{(i)}, {\treetilde\mu}_3^{(i)} \right) | N_{{\rm ring}}^{(i)}, {\treetilde\mu}_3^{(i)} \right) \, \right] \, .
\end{equation}
\item
Generate a large number (here, we have chosen $20'000$) of ``artificial'' arrays as Eq.~\eqref{eq:DefineDataSetForLogLikelihood}, where each entry is numerically sampled from $p(N_{\rm tree}, N_3 | N_{\rm ring}, {\treetilde\mu}_3)$ and, for each of them, calculate the corresponding log-likelihood as in Eq.~\eqref{eq:logL}.
\item
Calculate the $p$-value of each of the $200$ observed arrays, defined as the fraction of artificial arrays whose log-likelihood is {\it smaller} than the likelihood of the observed array.
Rank these $200$ $p$-values, from the smallest to the largest.
\item
Compute the empirical cumulative fraction ($f(p)$) of ranked $p$-values {\it smaller} than some $x\in[0,1]$.
\end{enumerate}

If Eq.~\eqref{eq:probability} of main text with the true parameters is indeed the correct data-generating model, then the resulting $p$-values should be approximately uniformly distributed in the interval $[0,1]$.
In that case, the empirical $f(p)$ should lie should lie approximately along the diagonal line $y=x$ while the observed average $p$-value, ${\bar p}_{\rm obs}$, should be close to $0.5$. 
As a complementary quantitive summary of this test, we also report the probability of observing an average $\bar p$ as far away from $0.5$ as ${\bar p}_{\rm obs}$.
For this we first use that when sampling individual numbers $X_i$ uniformly on the interval $[0,1]$, the expectation value and variance are found as $\mathbb{E}(X_i)=0.5$ and ${\rm Var}(X_i)=\frac{1}{12}$.
For a sample of size $200$, the expectation value and variance of the average ${\bar X}$ follow as $\mathbb{E}({\bar X})=0.5$ and ${\rm Var}(X)=\frac{1}{12 \cdot 200} =\frac{1}{2400}$.
By the central limit theorem, the distribution of $\bar X$ is approximately normal: ${\mathcal N}\!\left(0.5, \sigma^2 =\frac{1}{2400}\right)$.
Therefore, we can approximate the probability of finding an average $\bar p$-value as far from the average as ${\bar p}_{\rm obs}$, as:
\begin{eqnarray}
P\!\left( | {\bar p}-0.5 | > | {\bar p}_{\rm obs} - 0.5 | \right) = 2\left( 1 - \Psi\left( \frac{| {\bar p}_{\rm obs} - 0.5 |}{\sigma} \right) \right) , \,\,\, {\rm where} \,\,\, \Psi(x) = \frac{1}{\sqrt{2\pi}} \int_{-\infty}^x e^{-t^2/2} dt \, .
\end{eqnarray}

The plots in Fig.~\ref{fig:pNtreeN3-Validation} shows that the uniformity property holds much better when Eq.~\eqref{eq:probability} of main text with the correct parameters is used than when false values of $(N_{\rm ring}, {\treetilde\mu}_3)$ are used. 
In particular, the obtained values of $P\left( | {\bar p} - 0.5| > |{\bar p}_{\rm obs} - 0.5 | \right)$ quantify that only with the correct parameters, Eq.~\eqref{eq:probability} of main text provides a proper description of the data compared with the misspecified alternatives.
%This indicates that the hypothesis encoded in Eq.~\eqref{eq:probability} of main text provides a better description of the data than the misspecified alternatives. %None of the resulting $p_{\rm logL}$-values is sufficiently small to reject Eq.~\eqref{eq:probability} as the underlying probability distribution of the sampled data. %We also performed a combined statistical test across all input-parameter sets. %In this case, we summed the log-likelihoods of all 20 individual datasets and compared that combined value against 20'000 artificial datasets of equal total size (4,000 datapoints). %Each artificial dataset was generated by using the appropriate parameter values in Eq.~\eqref{eq:probability} for each block of 200 datapoints. %The resulting combined $p_{\rm logL}$-value was 0.6, providing further confirmation of our exact expressions.

%%%
\section{Asymptotic behaviours}\label{sec:Asymptotics}
%%%

%%%
\subsection{Partition function for tightly double-folded rings, $Z_{\rm ring}$}\label{sec:AsymptoticsZring}
%%%
The expression in Eq.~\eqref{eq:Z_ring} of main text is exact and can, in principle, be used to compute the expectation value of any observable $\langle \mathcal{A} \rangle$, provided one knows how $\mathcal{A}$ depends on the number of branch points $N_3$ and the tree size $N_{\rm tree}$.
However, for large $N_{\rm tree}$, factorial terms become numerically cumbersome. To address this, we derive analytical expressions in the limit $N_{\rm tree} \to \infty$, which are not only easier to evaluate but can also appear more intuitive for applications to realistic physical scenarios.
%
%\subsection{Scaling of the partition function}
%

The partition function $Z_{\rm ring}(N_{\rm tree}, {\treetilde\mu}_3)$ (Eq.~\eqref{eq:Z_ring} of main text) can be approximated using a scaling relation.
To do so, we first take the continuum limit and replace the sum over $N_3$ with the integral:
\begin{equation}\label{eq:Zint_1}
\frac{Z_{\rm ring}(N_{\rm tree}, {\treetilde\mu}_3)}{c^{\,N_{\rm tree}-1}} \approx 2\!\int_{0}^{N_{3,{\rm max}}} \frac{(N_{\rm tree}-1)!}{(N_3+2)! \, (N_{\rm tree}-2N_3-2)! \, N_3!} \, \exp\!\left[ \beta {\treetilde\mu}_3 N_3 \right] \, dN_3 \, ,
\end{equation}
which, after introducing the scaled variable $x = (N_3 + 1) / N_{\rm tree}$, becomes
\begin{equation}\label{eq:Zint_2}
\frac{Z_{\rm ring}(N_{\rm tree}, {\treetilde\mu}_3)}{c^{N_{\rm tree}-1}} \approx 2 \exp[-\beta {\treetilde\mu}_3] \, \int^{\frac{N_{3{\rm, max}}+1}{N_{\rm tree}}}_{\frac{1}{N_{\rm tree}}} \frac{N_{\rm tree}! \,\exp{\left[\beta {\treetilde\mu}_3 x N_{\rm tree} \right ]}}{\left(N_{\rm tree}\left(x+\frac{1}{N_{\rm tree}}\right)\right)!\, \left(N_{\rm tree}\left(1-2 x\right)\right)!\,\left(N_{\rm tree}\left(x-\frac{1}{N_{\rm tree}}\right)\right)!} \, dx \, .
\end{equation}
Applying Stirling's approximation ($n! \sim \sqrt{2\pi n} (n/e)^n$) and subsequently neglecting terms of order $\mathcal{O}(1/N_{\rm tree})$, we obtain
\begin{eqnarray}\label{eq:Zint_3}
\frac{Z_{\rm ring}(N_{\rm tree}, {\treetilde\mu}_3)}{c^{N_{\rm tree}-1}} 
& \approx &
\frac{\exp[-\beta {\treetilde\mu}_3]}{\pi N_{\rm tree}} \int_{\frac1{N_{\rm tree}}}^{\frac{N_{3,{\rm max}}+1}{N_{\rm tree}}} dx \, \frac1{\sqrt{(1-2x)\left(x^2 - \frac1{N_{\rm tree}^2}\right)}} \exp[\beta {\treetilde\mu}_3 x N_{\rm tree}] \nonumber
\\
& & 
\times
\frac{
\left(\frac{1}{1-2x}\right)^{N_{\rm tree}}
\left(N_{\rm tree} (1-2x)\right)^{2 x N_{\rm tree}}}
{\left(N_{\rm tree} \left(x + \frac{1}{N_{\rm tree}}\right)\right)^{\left(x + \frac{1}{N_{\rm tree}}\right) N_{\rm tree}} 
\left(N_{\rm tree} \left(x - \frac{1}{N_{\rm tree}}\right)\right)^{\left(x - \frac{1}{N_{\rm tree}}\right) N_{\rm tree}}} \nonumber\\ %[1ex]
\nonumber\\
& \approx &
\frac{\exp[-\beta {\treetilde\mu}_3]}{\pi N_{\rm tree}} 
\int_{\frac{1}{N_{\rm tree}}}^{\frac{N_{3,{\rm max}}+1}{N_{\rm tree}}} 
\frac{1}{x \sqrt{1-2x}} 
\left(\frac{1}{1-2x}\right)^{N_{\rm tree}} 
\left(\frac{1-2x}{x}\right)^{2 x N_{\rm tree}} 
\exp[\beta {\treetilde\mu}_3 x N_{\rm tree}] \, dx
\end{eqnarray}
In the limit $N_{\rm tree} \to \infty$, this integral can be evaluated using the Laplace method:
\begin{equation}\label{eq:laplacemethod}
\int_a^b h(x)\, e^{N g(x)} \, dx \approx \sqrt{\frac{2\pi}{N\,|g''(x_0)|}} \, h(x_0)\, e^{N g(x_0)} \, , \,\,\,\,\,\, \mbox{for $N\rightarrow\infty$}
\end{equation}
where $x_0$ is the value for which $g'(x_0)=0$.
From Eq.~\eqref{eq:Zint_3}, we identify: 
\begin{eqnarray}
g(x) & = & -\log \left(1-2x\right) + 2x \log\left( \frac{1}{x}-2\right) + \beta {\treetilde\mu}_3 x \\
g'(x) & = & 2 \log\left(\frac{1}{x}-2\right) + \beta {\treetilde\mu}_3 \label{eq:gprime} \\
g''(x) & = & \frac{2}{ x\left(2x-1\right)} \\
h(x) & = & \frac{1}{x\sqrt{1-2x}}
\end{eqnarray}
Solving $g'(x_0)=0$ gives:
\begin{equation}
x_0 = \frac1{2 + \exp\left( -\beta {\treetilde\mu}_3 / 2 \right)} \, ,
\end{equation}
which is equal to the asymptotic mean fraction of branch points $\lambda({\treetilde\mu}_3)$ (see Eqs.~\eqref{eq:lambda} and~\eqref{eq:AsymptoticN1N3} of main text).
Note also that this result is entirely consistent with the ``branching probability'' $\lambda$ of Eq.~(14) of Ref.~\cite{vanderHoek2025} given a shift equal to $-\log(2)$ in the branching chemical potential following from Eq.~\eqref{eq:degeneracy of ring labels} of main text.
Finally, after simple manipulations, the resulting asymptotic form of the partition function is given, up to unimportant prefactors, by the simple expression:
\begin{equation}\label{eq:Zringscaling} 
Z_{\rm ring}(N_{\rm tree}, {\treetilde\mu}_3)
\approx \frac{ \exp\left(-\beta {\treetilde\mu}_3 \right)}{\sqrt{\pi\ x_0}} \, \frac{\exp\left( N_{\rm tree}\, g(x_0)\right)c^{N_{\rm tree}-1}}{N_{\rm tree}^{3/2}}
%& \approx & \frac{\exp\left(-\beta {\treetilde\mu}_3 \right)}c \, \sqrt{\frac{2+\exp \left(-\frac{\beta {\treetilde\mu}_3}2 \right)}{\pi}} \, \frac{\left(\left(1+2\exp\left(\frac{\beta {\treetilde\mu}_3}2 \right)\right) c\right)^{N_{\rm tree}}}{N_{\rm tree}^{3/2}} \, ,
\sim \frac{\left(\frac{c}{1-2\lambda}\right)^{N_{\rm tree}}}{N_{\rm tree}^{3/2}} \, ,
\end{equation}
whose exponential factor coincides with Eq.~\eqref{eq:AsymptoticZring} of main text.
%This agrees with Eq.~(16) of \cite{vanderHoek2025} up to the number of distinct wrappings of a single tree (Eq.~\eqref{eq:degeneracy of ring labels}).
%The asymptotic result is valid as long as $x_0$ is not too close to the integration bounds, i.e., for moderate ${\treetilde\mu}_3$, avoiding linear chains or maximally branched trees.

%%%
\subsection{Approximate mean tree size with reptons, $\langle N_{\rm tree}\rangle$}\label{sec:expectedtreesize}
%%%
When $N_{\rm tree}$ is not conserved, we can use the scaling relation Eq.~\eqref{eq:Zringscaling} and the central limit theorem to approximate the expected tree size $\langle N_{\rm tree} \rangle$ by extremizing $\log \left[ \Omega_{\rm rep}(N_{\rm ring}, N_{\rm tree}) \, Z_{\rm ring}(N_{\rm tree}, {\treetilde\mu}_3) \right]$ in Eq.~\eqref{eq:ElasticModel-Z} of main text.

Applying Stirling's approximation, we write to leading order:
\begin{eqnarray}\label{eq:LogOmegaRepZringStirling}
\log \left[ \Omega_{\rm rep}(N_{\rm ring}, N_{\rm tree}) \, Z_{\rm ring}(N_{\rm ring}, {\treetilde\mu}_3) \right]
& \propto & 
\kappa\!\left(2(N_{\rm tree}-1)+N_{\rm rep}-N_{\rm ring}\right) \nonumber\\
& & + N_{\rm ring}\log(N_{\rm ring}) - 2(N_{\rm tree}-1)\log\!\left(2(N_{\rm tree}-1)\right) \nonumber\\
& & - N_{\rm rep}\log(N_{\rm rep}) + N_{\rm tree} \log\!\left( \frac{c}{1-2\lambda} \right) \, .
\end{eqnarray}
with the Lagrange multiplier $\kappa$ enforcing the constraint Eq.~\eqref{eq:repton_constraint}.
Extremizing Eq.~\eqref{eq:LogOmegaRepZringStirling} with respect to $N_{\rm rep}$, $N_{\rm tree}$ and $\kappa$ yields
\begin{eqnarray}
\kappa & = & 1 + \log(N_{\rm rep}) \, , \label{eq:LogOmegaRepZringStirling-extr1} \\
2 \kappa & = &  2 \log\left(2(N_{\rm tree} -1)\right) + 2 - \log\!\left( \frac{c}{1-2\lambda} \right) \, , \label{eq:LogOmegaRepZringStirling-extr2} \\
N_{\rm ring} & = & 2\left(N_{\rm tree}-1\right)+N_{\rm rep} \, . \label{eq:LogOmegaRepZringStirling-extr3}
\end{eqnarray}
Solving Eqs.~\eqref{eq:LogOmegaRepZringStirling-extr1}-\eqref{eq:LogOmegaRepZringStirling-extr3} gives the asymptotic expression:
\begin{align}\label{eq:N_rep}
\frac{\langle N_{\rm tree}\rangle - 1}{N_{\rm ring}/2} = 1-\frac{\langle N_{\rm rep}\rangle/2}{N_{\rm ring}/2}
%\approx \frac{\sqrt{\left(1 + 2 \exp \left( \frac{\beta {\treetilde\mu}_3}{2}\right) \right) c}}{1+\sqrt{\left(1 + 2 \exp \left( \frac{\beta {\treetilde\mu}_3}{2}\right) \right) c}}
\approx \left( \sqrt{\frac{1-2\lambda}c} + 1 \right)^{-1} \, , 
\end{align}
which coincides with Eq.~\eqref{eq:AsymptoticNtree} of main text.
By a similar procedure, one gets the asymptotic expressions Eqs.~\eqref{eq:AsymptoticN1N3} and~\eqref{eq:AsymptoticN2} of main text for $\langle N_1\rangle$, $\langle N_2\rangle$ and $\langle N_3\rangle$.

%%%
\subsection{Repton distribution according to tree nodes' functionality: comparison to computer simulations}\label{sec:ReptonDistrPerNode}
%%%
In the Supporting Information of Ref.~\cite{Ghobadpour2025} the distribution of the number of reptons per tree node was discussed.  
Here we evaluate that distribution analytically using the formalism that reptons are randomly distributed over the ring.  

%%%
\begin{figure*}%[ht]
\includegraphics[width=0.75\textwidth]{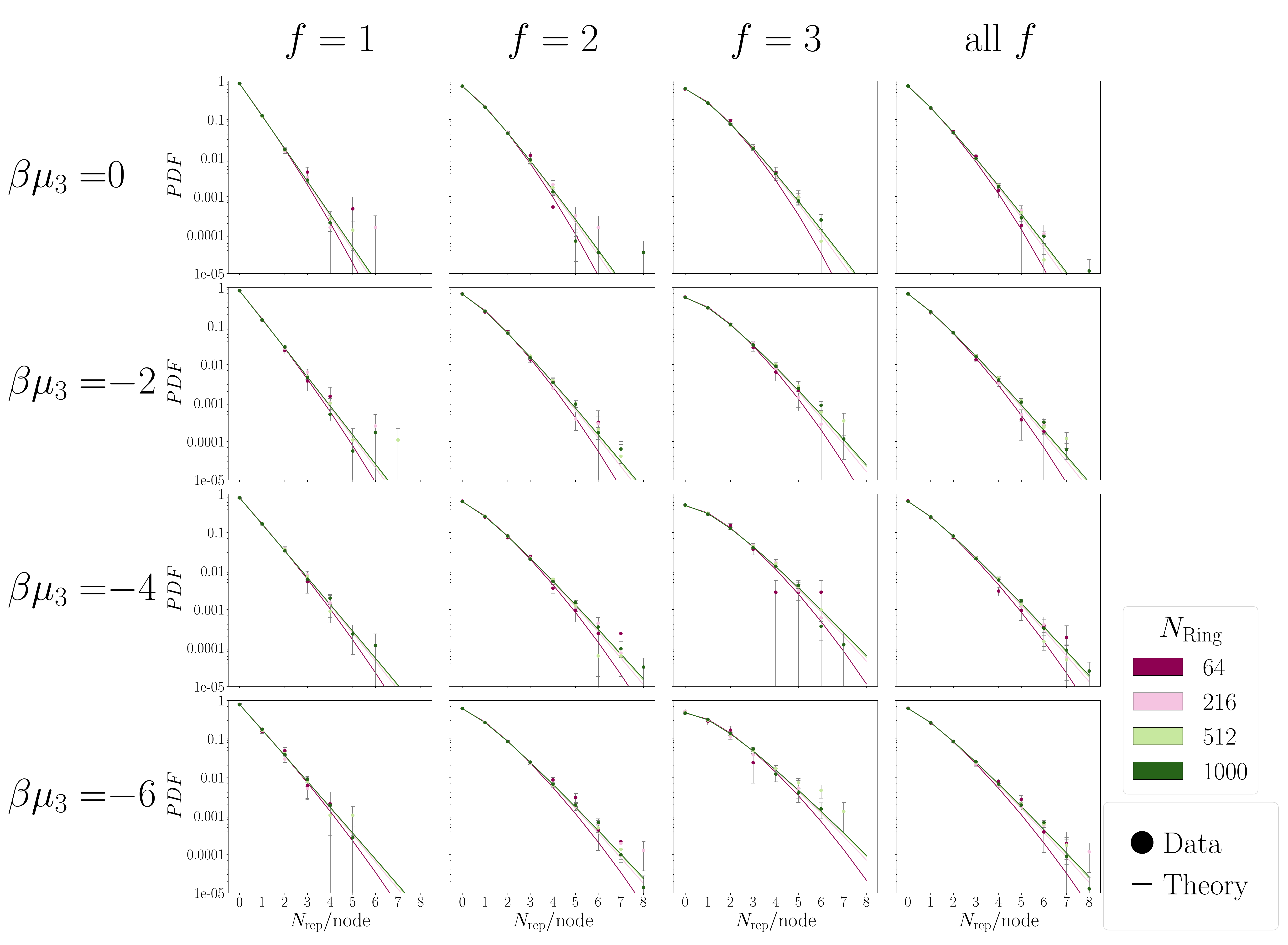}
\caption{
Probability distribution functions (PDF) for the number of reptons on tree nodes of different functionality $f$, for given $N_{\rm ring}$ and for different values of the branching chemical potential $\beta {\treetilde\mu}_3$ (see legends).
Symbols are for PDF's obtained from simulation data of the elastic lattice model and lines are for the theoretical prediction Eqs.~\eqref{eq:repton_distribution} and~\eqref{eq:repton_distribution_allf}.
In the theoretical formulas, $N_{\rm tree}$, $N_{\rm rep}=N_{\rm ring}-2(N_{\rm tree}-1)$ and $N_{f_i}$ have been replaced by, respectively, the asymptotic mean values Eq.~\eqref{eq:AsymptoticNtree}, Eq.~\eqref{eq:AsymptoticN1N3} and Eq.~\eqref{eq:AsymptoticN2} of main text.
}
\label{fig:reptondistributions}
\end{figure*}
%%%

We start from a tree of size $N_{\rm tree}$ around which a minimal ring of size $2(N_{\rm tree}-1) $ monomers is wrapped. 
By construction of the wrapping procedure, each tree-node $i$ with functionality $f_i$ has exactly $f_i$ ring monomers on it before placing any reptons. 
Reptons are then placed uniformly at random on the ring; hence the probability to place a repton on a particular tree-node $i$ is proportional to the number of ring-monomers already located on that node. 
As a consequence, the probability of choosing a given tree-node gets altered in the course of sequentially placing the reptons on the ring.
This growth-with-reinforcement process is described by the so-called Polya urn model~\cite{PolyaUrnModel}.
According to this model, after all reptons have been placed on the ring the probability of finding $k$ reptons on a tree node $i$ with functionality $f_i$ is given by the beta-binomial probability distribution:
\begin{equation}\label{eq:repton_distribution}
P(k | f_i, N_{\rm tree}, N_{\rm rep}) = \binom{N_{\rm rep}}{k} \frac{{\rm B}\!\left[k+f_i,N_{\rm rep} -k + 2\left(N_{\rm tree} -(f_i + 1)\right)\right]}{{\rm B}\left[f_i, 2\left(N_{\rm tree} -(f_i + 1)\right)\right]} \, ,
\end{equation}
where ${\rm B}\!\left[\alpha, \beta \right]=\frac{\Gamma(\alpha)\Gamma(\beta)}{\Gamma(\alpha +\beta)}$ is the so-called beta-function and $\Gamma(x)$ is the standard gamma-function.
Then, from Eq.~\ref{eq:repton_distribution} one can compute the distribution of reptons over tree nodes, regardless of their specific functionality, by just averaging the $f$-dependent distribution with the fraction $N_f / N_{\rm tree}$ of $f$-functional nodes, {\it i.e.}
\begin{equation}\label{eq:repton_distribution_allf}
P(k | N_{\rm tree}, N_{\rm rep}, N_3) = \sum^{f_i=3}_{f_i=1} \frac{N_{f_i}}{N_{\rm tree}} P(k | f_i, N_{\rm tree}, N_{\rm rep}) \, .
\end{equation}
To validate Eqs.~\eqref{eq:repton_distribution} and~\eqref{eq:repton_distribution_allf}, we have evaluated numerically repton distributions using again the elastic lattice model for tightly double-folded rings~\cite{Ghobadpour2021,Ghobadpour2025}, for $N_{\rm ring} \in \{ 64,216,512,1000 \}$ and branching chemical potential $\beta {\treetilde\mu}_3 \in \{ 0, -2, -4, -6 \}$.
As shown in Fig.~\ref{fig:reptondistributions}, the formulas~\eqref{eq:repton_distribution} and~\eqref{eq:repton_distribution_allf} reproduce the data very accurately (lines {\it vs.} symbols).

%%%
\begin{figure*}%[ht]
\includegraphics[width=0.75\textwidth]{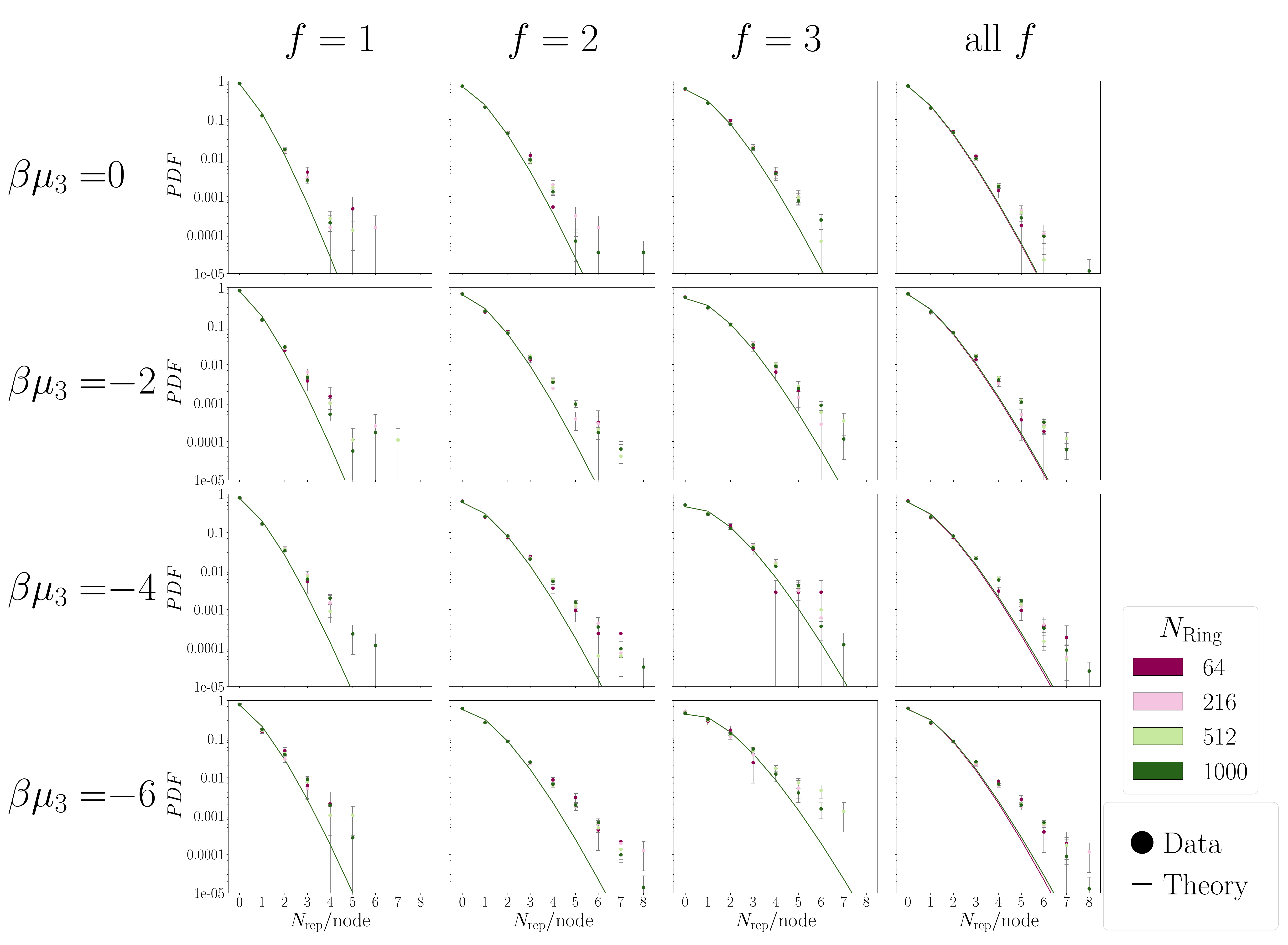}
\caption{
Comparison between the same data shown in Fig.~\ref{fig:reptondistributions} (symbols) and the theoretical distributions for reptons (lines) following Poisson law (Eqs.~\eqref{eq:repton_distribution_poisson} and~\eqref{eq:repton_distribution_allf}).
}
\label{fig:reptondistributions_poisson}
\end{figure*}
%%%

For comparison, the distribution of reptons proposed in~\cite{Ghobadpour2025} was not Eq.~\eqref{eq:repton_distribution}, but the following Poisson-like function:
\begin{equation}\label{eq:repton_distribution_poisson}
P(k | f_i, N_{\rm tree}, N_{\rm rep})= \frac{\lambda_{f_i}^k \exp(-\lambda_{f_i})}{k!} \, ,
\end{equation}
with $\lambda_{f_i} =\frac{N_{\rm rep}f_i}{2 (N_{\rm tree}-1)}$.
As displayed in Fig.~\ref{fig:reptondistributions_poisson}, Eq.~\eqref{eq:repton_distribution_poisson} does not capture well the behavior of simulated rings.

%%%
\section{Trees of arbitrary functionality}\label{sec:arbitrary_functionality}
%%%
To reinforce the claim made in Eq.~\eqref{eq:Multiplicity_with_f>3} of main text, we use again the Supporting Information of~\cite{Ghobadpour2025}.
There, several ensembles of tightly double-folded rings were simulated at fixed ring size $N_{\rm ring}=216$ and with no chemical potential applied, but with different imposed maximum functionalities $f_{\rm max} \in \{ 3,6,9,12 \}$.
In these ensembles the usual tree constraints (Eqs.~\eqref{eq:n1_res} and~\eqref{eq:n2_res} of main text) generalize to
\begin{eqnarray}
\sum^{f=f_{\rm max}}_{f=1}N_f & = & N \, , \label{eq:generalconstraintN} \\
\sum^{f=f_{\rm max}}_{f=3} (f-2) N_f & = & N_1 - 2 \, . \label{eq:generalconstraintN1}
\end{eqnarray}
For generality, in the following we associate a functionality-specific chemical potential ${\treetilde\mu}_f$ to each node of functionality $f$. %, generalizing the Hamiltonian (Eq.~\eqref{eq:BranchingHam}):
%
%\begin{equation}\label{eq:BranchingHam_general}
%{\mathcal H}_{\rm  gen}= \sum^{f=f_{\rm max}}_{f=1} {\treetilde\mu}_f N_f \, .
%\end{equation}
%%

The asymptotic expression for the mean total number, $\langle N_f\rangle$, of nodes of functionality $f$ can be computed using the same methods introduced in Sec.~\ref{sec:expectedtreesize}.
As before, we invoke the central limit theorem and extremize $\log\left[ \Omega_{\rm ring}(N_{\rm tree}, \{ N_f \}) \, e^{\sum_{f=1}^{f_{\rm max}} \beta \mu_f N_f} \right]$, with $ \Omega_{\rm ring}(N_{\rm tree}, \{ N_f \})$ given by Eq.~\eqref{eq:Multiplicity_with_f>3} in main text.
Incorporating the constraints~\eqref{eq:generalconstraintN} and~\eqref{eq:generalconstraintN1} using Lagrange multipliers $\kappa_1$ and $\kappa_2$, we obtain to leading order the expression:
\begin{eqnarray}
\log\left( \Omega_{\rm ring}(N_{\rm tree},\{N_f\}) \right) + \sum_{f=1}^{f_{\rm max}} \beta \mu_f N_f
& \approx &
\sum_{f=1}^{f_{\rm max}} \beta {\treetilde\mu}_f N_f + (N_{\rm tree}-1)\log(N_{\rm tree}-1) - \sum_{f=1}^{f_{\rm max}} N_f \log N_f \nonumber \\
& & + \kappa_1 \left( N_{\rm tree}-\sum_f N_f \right) + \kappa_2 \left( N_1 - 2 -\sum^{f= f_{\rm max}}_{f=3} (f-2) N_f \right) \, .
\end{eqnarray}
Extremizing with respect to $N_f$, $\kappa_1$ and $\kappa_2$ yields the system
\begin{eqnarray}
N_f & = & \exp\left(-(f-2)\kappa_2 - \kappa_1 -1 + \beta {\treetilde\mu}_f \right) \, , \\
N_1 & = & \sum^{f= f_{\rm max}}_{f=3} (f-2) N_f +2 \, , \\
N_{\rm tree} & = & \sum_{f=1}^{f_{\rm max}} N_f \, ,
\end{eqnarray}
which, after neglecting terms ${\mathcal O}(1/N_{\rm tree})$ and smaller, can be written in the compact form:
\begin{eqnarray}
\frac{N_f}{N_{\rm tree}}
& = & \frac{\exp\left(-(f-2)\kappa_2 + \beta {\treetilde\mu}_f \right)}{\sum^{f'= f_{\rm max}}_{f'=1} \exp\left(-(f'-2)\kappa_2 + \beta {\treetilde\mu}_{f'} \right)} \, , \label{eq:arbitrary_f_a} \\
\frac{N_1}{N_{\rm tree}}
& = & \sum^{f= f_{\rm max}}_{f=3} (f-2) \frac{N_f}{N_{\rm tree} } \, . \label{eq:constrainedendpoints}
\end{eqnarray}
%
%Here, in the second line, we took the large $N_{\rm tree}$-limit.
Then, introducing $\alpha=e^{-\kappa_2}$ Eq.~\eqref{eq:constrainedendpoints} is reduced to the simple polynomial equation:
\begin{equation}\label{eq:alphapolynomial_gen}
1 = \sum_{f=3}^{f = f_{\rm max}} (f-2) \, \alpha^{f-1} \exp[ -\beta( {\treetilde\mu}_1 - {\treetilde\mu}_f ) ] \, ,
\end{equation}
which must be solved numerically to find the optimal $\alpha$.
After that, from Eq.~\eqref{eq:arbitrary_f_a} one obtains 
\begin{equation}\label{eq:expected_nf_gen}
\frac{N_f}{N_{\rm tree}} = \frac{\alpha^f \exp(\beta {\treetilde\mu}_f)}{\sum_{f'=1}^{f'= f_{\rm max}} \alpha^{f'} \exp(\beta {\treetilde\mu}_{f'})} \, .
\end{equation}
Notice that a closely related expression in the context of a mean-field approximation for {\it interacting} ({\it i.e.}, non-ideal) trees was derived recently in Ref.~\cite{Marcato2025}.
For branching chemical potentials $\{ {\treetilde\mu}_f = 0 \}$, it is easy then to check that Eqs.~\eqref{eq:alphapolynomial_gen} and~\eqref{eq:expected_nf_gen} give Eq.~\eqref{eq:muf>3-LimitingCases} of main text for the two limiting cases $f_{\rm max}=3$ and $f_{\rm max}=\infty$. %we obtain Eq.~\eqref{eq:alphapolynomial} from Eq.~\eqref{eq:alphapolynomial_gen} and Eq.~\eqref{eq:expected_nf} from  Eq.~\eqref{eq:expected_nf_gen}.

%%%
%\end{appendix}
%%%

\end{document}